\documentclass[a4paper,superscriptaddress,floatfix]{revtex4}

\usepackage{epsfig}
\usepackage{amsfonts}
\usepackage{amsmath}
\usepackage{mathrsfs}
\usepackage{amssymb}
\usepackage{amsthm}
\usepackage{float}
\usepackage{array}
\usepackage{booktabs}
\usepackage{color}
\usepackage{graphicx}
\usepackage{hyperref}
\usepackage{multirow}
\usepackage{slashed}
\usepackage{subfigure}
\usepackage{theorem}
\usepackage{textcomp}

\newcommand\numberthis{\addtocounter{equation}{1}\tag{\theequation}}

\allowdisplaybreaks[1]

\numberwithin{equation}{section}

\begin{document}

\title{Constrained Vector-like quark model with perturbative unitarity}

\author{Elena Accomando}
\email[E-mail: ]{e.accomando@soton.ac.uk}
\affiliation{School of Physics \& Astronomy, University of Southampton,
        Highfield, Southampton SO17 1BJ, UK}
\affiliation{Particle Physics Department, Rutherford Appleton Laboratory, 
       Chilton, Didcot, Oxon OX11 0QX, UK}

\author{Jake Brannigan}
\email[E-mail: ]{jlb3g17@soton.ac.uk}
\affiliation{School of Physics \& Astronomy, University of Southampton,
        Highfield, Southampton SO17 1BJ, UK}

\author{Jacob Gunn}
\email[E-mail: ]{jg8g17@soton.ac.uk}
\affiliation{School of Physics \& Astronomy, University of Southampton,
        Highfield, Southampton SO17 1BJ, UK}

\author{Yuheng Huyan}
\email[E-mail: ]{yh5g17@soton.ac.uk}
\affiliation{School of Physics \& Astronomy, University of Southampton,
        Highfield, Southampton SO17 1BJ, UK}

\author{Sam Mulligan}
\email[E-mail: ]{sm25g16@soton.ac.uk}
\affiliation{School of Physics \& Astronomy, University of Southampton,
        Highfield, Southampton SO17 1BJ, UK}

\begin{abstract}
\noindent
{We propose a Vector-like quark (VLQ) model constrained by the requirement of perturbative unitarity. In this scenario, the neutral and charged couplings of the new heavy VLQs to SM quarks and bosons are strictly related. We derive the corresponding sum rules and display the definite structure of the couplings. We analyse, qualitatively, the phenomenological consequences of this model in terms of expected width, production mechanism and decay channels. We show that the upcoming LHC run 3 could have sensitivity to such particles but the search strategy should be adapted to cover wide resonances decaying mainly in the charged channel $T(B)\rightarrow Wq$.
}
\end{abstract}

\maketitle


\setcounter{footnote}{0}

\section{Introduction}

Though the Standard Model (SM) has proved itself amazingly successful, well established tensions of the SM with precision measurements indicate the need for new physics. The CKM matrix encodes couplings for each of the three generation quarks. By construction, within the SM the CKM matrix is unitarity. For the first row, unitarity relations predict that 
$|V_{ud}|^2+|V_{us}|^2+|V_{ub}|^2=1=C^1_3$. Yet, the current best experimental determinations of $V_{ud}$ and $V_{us}$ find $C^1_3=0.99798\pm 0.0003$ that is a disagreement greater than 4$\sigma$ \cite{Belfatto:2019swo}. The recent dataset collected after 2018 \cite{FlavourLatticeAveragingGroup:2019iem} disfavours the CKM unitarity for three generations of quarks at 99.998\% C.L. Such a deviation from the CKM unitarity seems to suggest the presence of new physics. The hypothetical VLQs could explain this misalignment between the SM predictions and observed data; this note will address this Beyond Standard Model (BSM) scenario. This class of models can indeed naturally lead to deviations from the CKM unitarity capable of accommodating the recent measurements of $|V_{us}|$ and $|V_{ud}|$ \cite{Belfatto:2019swo, Branco:2021vhs}.

\noindent
Vector-like quarks (VLQs) are spin 1/2 particles characterised by having the left- and right-handed components defined by the same colour and electroweak quantum numbers. Under a given gauge group, their left and right-handed projections belong to the same representation. They naturally emerge in many BSM scenarios. VLQs with a TeV-scale mass are indeed strongly motivated by two main theoretical frameworks: models where the Higgs is a pseudo-Goldstone boson \cite{Perelstein:2003wd, Contino:2006qr, Matsedonskyi:2012ym} and partial compositeness flavour theories \cite{Contino:2006nn}. In the first case, vector-like quarks are required to induce the electroweak symmetry breaking and explain the lightness of the observed Higgs boson. In the latter, they are part of the predicted extra heavy fermion resonances. Explicit realisations of these two basic mechanisms are little Higgs models, composite Higgs models, and their holographic versions. VLQs also appear in extra dimension theories with quarks propagating in the bulk, and in grand unified or string $E_6$-models. There is a vast literature on the development of general VLQ models realised by extending the SM with the sole inclusion of VLQs coming in three possible representations i.e. singlet, doublet and triplet. For a review see Refs.~\cite{Cacciapaglia:2010vn,Cacciapaglia:2011fx,Cacciapaglia:2018lld,Botella:2012ju}. 

\noindent
From the experimental point of view, vector-like quarks are the only coloured fermions still allowed by the most recent data. The forth generation of quarks with chiral couplings is indeed highly disfavoured by the electroweak precision tests combined with Higgs data \cite{Eberhardt:2012gv, Djouadi:2012ae}. Vector-like quarks can evade such exclusion bounds as they are not chiral, a priori, and do not have  to acquire their mass via the Higgs mechanism. Therefore, VLQs are receiving a lot of attention at the LHC. 
The CERN CMS collaboration has set mass exclusion bounds up to 1650 ${\rm GeV}$, depending on the vector-like quark type, coupling and decay width \cite{CMS:2018dcw}. The CERN ATLAS collaboration has set analogous exclusion limits at 1030 (1210) ${\rm GeV}$ for singlet (doublet) up-type VLQ and 1010 (1140) ${\rm GeV}$ for singlet (doublet) down-type VLQ \cite{Vale:2018bpf}.

\noindent
Direct experimental searches for such heavy states assume that a VLQ resonance can be described by a Breit-Wigner (BW) line-shape, standing over the SM background, when looking at the cross section distribution in the invariant mass of the VLQ decay products. Hence, the new physics signal is expected to have a peaking structure, concentrated in an interval centred around its mass. The VLQs existence could also be inferred from indirect searches. These new particles could in fact give rise to potentially striking effects in low energy physics. New tree-level flavour-changing neutral currents (FCNC), mediated by the weak $Z$-boson and/or the Higgs boson, could appear. VLQs could also introduce new sources of CP violation. The landscape of signatures is quite varied and interesting, a priori. 

\noindent
There is a strong interplay between direct and indirect searches for VLQs. The mixing between the SM quarks and VLQs, which generates the FCNC, has an impact both on the production process and the open decay channels of these latter particles. This result, in turn, affects the way the new heavy particles are searched for at the LHC. Direct experimental searches are typically optimised for narrow resonances and, generally, assume a pure BW line-shape over the SM background, when looking at the invariant mass of the VLQs decay products. On the basis of this assumption, in absence of evidence, the 95\% Confidence Level (CL) upper bound on the cross section is derived and limits on the mass and couplings of the resonance are extracted within some benchmark model. In this approach, theoretical cross sections are usually calculated in Narrow Width Approximation (NWA); more rarely, they might include Finite Width (FW) and interference effects. The main production channel of VLQs at the LHC is the pair production, mediated by the gluon.
The extra heavy quarks subsequently decay into SM particles, namely ordinary quarks plus a Higgs or a gauge boson, with branching ratios that are mostly determined by their gauge quantum numbers. These decays occur through the mixing of the new quarks with the SM ones. The very same mixing opens up the possibility to singly produce the new quarks. This becomes the dominant production mechanism for high enough masses.

\noindent
The purpose of this paper is building a VLQ model that preserves the perturbative unitarity of the theory, explicitly. The inclusion of this property constrains the VLQs couplings to the SM gauge bosons and matter, enhancing the predictivity of the model. The advantage of this approach is twofold. Firstly, one can find relations between the charged current and neutral current couplings. Secondly, these relations can be used to extract bounds on such couplings from different experiments. We consider two possible group representations for the VLQs: singlet, $T$ or $B$, and doublet $(T, B)$. To model the VLQs couplings to SM particles, by imposing perturbative unitarity, we then use candle processes involving SM vector bosons and quarks along with at least one vector-like quark.  

\noindent 
The format of this note is as follows. In Sect.~\ref{sec:lagrangian}, we define the VLQs representations considered in this paper and the corresponding Lagrangian for the charged and neutral currents. In Sect.~\ref{sec:pu}, we derive the relations between charged and neutral VLQs couplings, by requiring perturbative unitarity,  and define our constrained model. In Sect.~\ref{sec:pheno}, we illustrate some phenomenological aspects of the vector-like quarks with constrained couplings, which could have significant consequences for their search at the LHC. Finally, we conclude in Sect.~\ref{sec:conc}.

\section{VLQs representations}
\label{sec:lagrangian}

\noindent
New heavy vector-like quarks, with renormalisable couplings to the SM gauge bosons and quarks, can only appear in seven possible multiplets characterised by well defined $SU(3)_c\times SU(2)_L\times U(1)_Y$ quantum numbers \cite{delAguila:2000aa, delAguila:2000rc}: $T$- and $B$-singlets, $(X, T)$-, $(T, B)$- and $(B, Y)$-doublets, $(X, T, B)$- and $(T, B, Y)$-triplets. The new heavy partners of the SM top and bottom quarks, $T$ and $B$, have electric charge $2/3$ and $-1/3$, respectively. The exotic vector-like quarks, $X$ and $Y$, have instead a charge different from the SM up-type and down-type quarks. They are indeed characterised by a fractional charge of $5/3$ and $-4/3$, respectively. 

\noindent
The Lagrangian for the charged current induced by the $W$-boson can be written as:
\begin{equation}
{\mathcal{L}}_W = -{g_W\over\sqrt{2}}\bar U\gamma^\mu (V^L_{UD}P_L + V^R_{UD}P_R)DW^+_\mu + H.c.
\end{equation}
where $P_L$ and $P_R$ are the left- and right-handed projectors, respectively.
The symbol $U(D)$ stands for any up(down)-type quarks, including the extra vector-like quark. The couplings $V^{L, R}_{UD}$ are the entries of the left-handed and right-handed CKM matrix, extended by the addition of the hypothetical vector-like quark(s). In this scenario, the SM CKM matrix appears as the $3\times 3$ block of the left-handed matrix (within the SM the right-handed entries are null).

\noindent
The Lagrangian for the neutral current induced by the weak $Z$-boson is:
\begin{equation}
{\mathcal{L}}_Z = -{g_W\over{c_W}}\bar q\gamma^\mu \left ((\pm c^{ZL}_{qq^\prime}P_L \pm c^{ZR}_{qq^\prime}P_R) - 2Q_qs^2_W\delta_{qq^\prime} \right )q^\prime Z_\mu + H.c.
\end{equation}
where $q$ and $q^\prime$ describe any quarks, including the heavy vector-like quark(s). Here, $s_W(c_W)$ is the Weinberg angle sine(cosine) and $Q_q$ the electric charge of the quark $q$. The plus sign is for up-type quarks, the minus for down-type quarks. The first two terms in the Lagrangian can give rise to flavour changing neutral currents, in case $q$ and $q^\prime$ were different. In that event, the complex parameters $c^{ZL}$ and $c^{ZR}$ represent the strength of the FCNC of the weak neutral $Z$-boson.

\noindent
Finally, the Lagrangian for the neutral current induced by the Higgs boson is:
\begin{equation}
{\mathcal{L}}_h = -{g_W\over{2M_W}}\bar q\gamma^\mu (\pm M_Lc^{hL}_{qq^\prime}P_L \pm M_R	c^{hR}_{qq^\prime}P_R)q^\prime H + H.c.
\end{equation}
where $M_{L, R}$ are generic masses, which will later be related to the masses of the $q$ and $q^\prime$ quarks. One can notice that, as for the $Z$-boson mediated neutral current, also in the case of the Higgs boson one can have the appearance of FCNC whose strength is represented by the complex parameters $c^{hL}$ and $c^{hR}$. Moreover, oppositely to the SM predictions, in these BSM scenarios the left- and right-handed couplings of the Higgs to fermions are not identical. This implies that the Higgs might have a pseudo-scalar component whose magnitude depends on the difference $(c^{hL}_{qq^\prime}-c^{hR}_{qq^\prime})/(c^{hL}_{qq^\prime}+c^{hR}_{qq^\prime})$.

\noindent
In this paper, we shall restrict ourselves to extensions of the SM with only one extra multiplet. In particular, as anticipated in the introduction, we analyse three types of multiplet: $T$-singlet, $B$-singlet and $(T, B)$-doublet. 

\section{Perturbative unitarity}
\label{sec:pu}

\noindent
In this section, we analyse the consequences of imposing the perturbative unitarity on the Standard Model extended by the inclusion of vector-like quarks. Our approach in this investigation is intended to be generic and model independent; therefore, we will not specify beforehand a representation with which we will derive our results. For the same reason, we do not make any assumptions on the couplings of the VLQs with SM particles. 

\noindent
We select as candle process the vector-boson-fermion scattering (VBFS)
\begin{equation}
W^+_L(k_1) d(p_1) \rightarrow W^+_L(k_2) b(p_2)
\end{equation} 
where $d$ and $b$ are the down- and bottom-quark, respectively, and $W^+$ the weak $W$-bosons. The symbol $L$ denotes the longitudinal polarisation of the vector bosons while  $k_1, p_1, k_2, p_2$ are the 4-momenta of the particles. We generalise the result in the next section. The choice of this channel is well motivated. When considering weak interactions, the preservation of perturbative unitarity is intimately tied to the Higgs mechanism and the Higgs boson. The Higgs mechanism is indeed responsible for the longitudinal polarisations of the weak vector bosons, whose charged version is present in our chosen interaction. As a thumb rule, for each weak gauge boson appearing as an external particle in a two-by-two scattering process, the matrix element acquires an energy growth proportional to the total Centre-of-Mass (CoM) energy, $\sqrt{s}=E^{CM}_{TOT}$. The Higgs boson is then required to cancel these energy dependent terms in the matrix element, if all the couplings involved have the correct structure. Our candle process (including its generalisations) is therefore excellent for extracting informations on the relations between the different fermion couplings to gauge bosons by using the requirement of perturbative unitarity. The above channel has in fact two $W$-bosons, giving rise to an increase with energy of the matrix element proportional to $s$, and two fermions allowing for the presence of extra vector-like quarks either as external particles in generalised processes or as a mediator.

\noindent
As already mentioned, we consider three group representations of the VLQs: $T$-singlet, $B$-singlet and $(T, B)$ doublet. From the requirement of perturbative unitarity, sum rules for the VLQs couplings with SM gauge bosons and quarks will be derived. The sum rules have implications on the phenomenology of the VLQs, which will be  discussed in the next section.

\begin{figure}[t!]
\begin{center}
\includegraphics[width=0.7\textwidth]{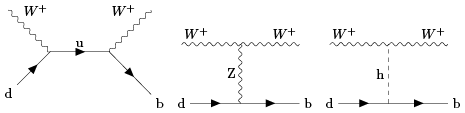}
\caption{Complete set of Feynman diagrams, at the tree level, contributing to the process $W^+ d\rightarrow W^+ b$. From left to right, the first graph is mediated by up-type quarks, the second and the third by the Z- and Higgs boson, respectively.}
\label{fig:FD_Sam}
\end{center}
\end{figure}

\noindent
At the tree level, there are three Feynman diagrams potentially contributing to this process. They are shown in Fig.~\ref{fig:FD_Sam}. The first of these diagrams is the only interaction of the three to be ordinarily permitted by the standard model. The remaining two diagrams represent flavour-changing neutral currents (FCNC), which can only happen at loop level in the SM and are highly suppressed by the GIM mechanism. As such, modification of SM observables by contributions from FCNCs could provide evidence for the presence of new vector-like quarks.

In order to analyse the behaviour of the matrix element at high energies and see what conditions on couplings and masses are required to guarantee perturbative unitarity, we consider longitudinally polarised $W$-bosons. We adopt the representation of the longitudinal polarisation vectors, $\epsilon^\mu_L$, given by the {\it Mathematica} package {\it Feyncalc} \cite{Shtabovenko:2020gxv, Shtabovenko:2016sxi}:
\begin{equation}
\epsilon^\mu_L(p_1)={{p_1^\mu*(k_1\cdot p_1 + k_2\cdot p_1)-(k_1+k_2)^\mu p_1^2}\over{\sqrt{p_1^2[(k_1\cdot p_1+k_2\cdot p_1)^2-p_1^2(k_1+k_2)^2}}}
\end{equation} 
with the 4-momentum assignment given above. Performing a Taylor expansion in the mass over energy ratio, we obtain the analytical expression for the matrix elements corresponding to the Feynman diagrams in Fig.~\ref{fig:FD_Sam}. The amplitude of the first graph mediated by the up-type quarks is given by:

\begin{align*} 
M_U &= 
\sum_{u}F_u\bar{u}_b(k_2)[A_uP_L + B_u P_R + C_u \slashed{p_1} P_R + D_u \slashed{p_1}P_L]u_d(p_2)
\end{align*}  
where the sum goes over all the possible up-type quarks: up, charm, top and the heavy $T$ vector like quark. The constants $A_u, B_u, C_u, D_u$ are functions of the CoM energy $s$ and masses:

\begin{align*} 
A_u &=
m_b (m_b^2 - M_W^2 - s)\left [V_{bu}^{L*}V_{ud}^L\left (s (s - M_W^2) + m_d^2 (m_d^2 - M_W^2 -2s)\right ) + V_{bu}^{L*}V_{ud}^R (-2m_um_dM_W^2)\right ] + \\
& (s - M_W^2 - m_b^2)\left [V_{bu}^{R*}V_{ud}^L m_u (s^2-M_W^2s-2m_d^2s-M_W^2m_d^2+m_d^4) -2m_dM_W^2s V_{bu}^{R*}V_{ud}^R\right ] 
\end{align*}   

\begin{align*} 
B_u &=
m_b (m_b^2 - M_W^2 - s)\left [V_{bu}^{R*}V_{ud}^R\left (s (s - M_W^2) + m_d^2 (m_d^2 - M_W^2 -2s)\right ) + V_{bu}^{R*}V_{ud}^L (-2m_um_dM_W^2)\right ] + \\
& (s - M_W^2 - m_b^2)\left [V_{bu}^{L*}V_{ud}^R m_u (s^2-M_W^2s-2m_d^2s-M_W^2m_d^2+m_d^4) -2m_dM_W^2s V_{bu}^{L*}V_{ud}^L\right ] 
\end{align*}   

\begin{align*} 
C_u &=
(s - M_W^2 - m_b^2)\left [V_{bu}^{R*}V_{ud}^R s(s-M_W^2-m_d^2) -m_um_d V_{bu}^{R*}V_{ud}^L (s+M_W^2-m_d^2)\right ] + \\
& m_b (m_b^2 - M_W^2 - s)\left [-m_d V_{bu}^{L*}V_{ud}^L(s + M_W^2 -m_d^2) + m_u V_{bu}^{L*}V_{ud}^R (s-M_W^2-m_d^2)\right ]
\end{align*}   

\begin{align*} 
D_u &=
(s - M_W^2 - m_b^2)\left [V_{bu}^{L*}V_{ud}^L s(s-M_W^2-m_d^2) -m_um_d V_{bu}^{L*}V_{ud}^R (s+M_W^2-m_d^2)\right ] + \\
& m_b (m_b^2 - M_W^2 - s)\left [-m_d V_{bu}^{R*}V_{ud}^R(s + M_W^2 -m_d^2) + m_u V_{bu}^{R*}V_{ud}^L (s-M_W^2-m_d^2)\right ]
\end{align*}   

\begin{equation} 
F_u = \left ({{-ig_W^2}\over{2 M_W^2}}\right ){1\over{s^3}}\left [ 1+{{(m_u^2+2M_W^2+m_b^2+m_d^2)}\over s}\right ].
\end{equation}   
In these formulae, $m_d$ and $m_b$ are the masses of the down- and bottom-quark, respectively. The mass of the up-type quark, mediator of the interaction, is indicated by $m_u$. 

\noindent
The amplitude of the second graph in Fig.~\ref{fig:FD_Sam}, mediated by the $Z$-boson, is given by:

\begin{align*} 
M_Z &= 
F_z{\bar u}_b(k_2)[A_zP_L + B_z P_R + C_z \slashed{p_1} P_R + D_z \slashed{p_1}P_L]
u_d(p_2) 
\end{align*}  
where
\begin{equation} 
A_z=m_bc^{ZL}_{bd}A+m_dc^{ZR}_{bd}B, ~~~~ B_z=m_bc^{ZR}_{bd}A+m_dc^{ZL}_{bd}B, ~~~~ C_z=c^{ZR}_{bd}C, ~~~~ D_z=c^{ZL}_{bd}C.
\end{equation}  
\noindent 
The functions $A, B, C$ and $F_z$ are defined as follows:

\noindent
\begin{align*} 
A &=
M_W^2\left [{t\over 2}(M_W^2-m_b^2)-(s-M_W^2)^2+m_b^2(m_b^2-M_W^2-s)\right ]+{{ts}\over 2}(2M_W^2-m_b^2+s) + \\
& m_d^2\left [M_W^2(-{t\over 2}+3s+3M_W^2-m_b^2-m_d^2)+{t\over 2}(m_b^2-s)\right ]
\end{align*}   

\noindent
\begin{align*} 
B &=
{t\over 2}M_W^2(7M_W^2-3m_b^2)+{{ts}\over 2}(6M_W^2+m_b^2-s)+M_W^2 \left [m_b^2(s+M_W^2)+(s-M_W^2)^2-m_b^4\right ] + \\
& m_d^2\left [{t\over 2}(s-3M_W^2-m_b^2)+M_W^2(m_d^2-3s-3M_W^2+m_b^2)\right ]
\end{align*}   

\noindent
\begin{align*} 
C &=
2M_W^2(s-M_W^2)^2-M_W^2m_b^2(t+2s+2M_W^2)+ts(m_b^2+2M_W^2)+t(3M_W^4-s^2)+\\
&m_d^2\left [2M_W^2(m_b^2-s-M_W^2)+t(s-m_b^2-M_W^2)\right ]
\end{align*} 
\noindent 
and 

\begin{equation} 
F_z =  \left ({{ig_W^2}\over{M_W^2}}\right ){1\over{s^2t}}\left [ 1+{{(2M_W^2+m_b^2+m_d^2)}\over s} +{M_z^2\over t}\right ]
\end{equation}  

\noindent
Finally, the diagram mediated by the Higgs boson is given by:
\begin{equation}
M_H = F_hA_h{\bar u}_b(k_2)[M_Lc^{hL}_{bd}P_L + M_Rc^{hR}_{bd}P_R]u_d(p_2)
\end{equation}
with
\begin{align*} 
A_h &= (s-M_W^2-m_d^2)[(s-M_W^2-m_b^2)(M_W^2-{t\over 2})- M_W^2(M_W^2+m_b^2-u)]\\
&-M_W^2(s-M_W^2-m_b^2)(M_W^2+m_d^2-u) + 2M_W^4(m_b^2+m_d^2-t)  
\end{align*}  
and
\begin{equation}
F_h={{ig_W^2}\over{2M_W^2}}{1\over{ts^2}}\left [1+{{2M_W^2+m_d^2+m_b^2}\over s}+{{M_h^2}\over t}\right ]
\end{equation}
where $M_h$ is the Higgs mass. In the above formulae, $s, t, u$ are the three Mandelstam variables defined as:
\begin{equation} 
s = (p_1 + p_2)^2, ~~~~ t = (p_1 - k_1)^2, ~~~~ u = (p_1 - k_2)^2.
\end{equation}   

\noindent
The modulus squared of the sum of the three amplitudes represented by the Feynman diagrams in Fig.~\ref{fig:FD_Sam}, $|M|^2=|M_U+M_Z+M_H|^2$, has been computed using the UNIX-based program FORM. Once expanded in terms of the Mandelstam variables, it displays a term proportional to the product $su$, which grows with the fourth power of the CoM energy $E_{CM}^4$, terms proportional to $t$ and $u$ increasing as $E_{CM}^2$, and terms constant or decreasing with increasing the energy. The highest order $su$-term comes from the modulus square of the sum of the two matrix elements mediated by the up-type quarks and the $Z$-boson, specifically, and it can be written as
\begin{equation}
I_{us}={g_W^4\over M_W^4}(-us)\left [\left |{1\over 2}\sum_u V_{bu}^{R*}V_{ud}^R+c^{ZR}_{bd}\right |^2 + \left |{1\over 2}\sum_u V_{bu}^{L*}V_{ud}^L+c^{ZL}_{bd}\right |^2\right ].
\end{equation}
\noindent
In order to restore perturbative unitarity at high energies, this term must be cancelled out. By imposing this requirement, a natural relation between the weak charged and neutral couplings can be extracted:
\begin{equation}
-{1\over 2}\sum_u V_{bu}^{R*}V_{ud}^R = c^{ZR}_{bd}; ~~~~ -{1\over 2}\sum_u V_{bu}^{L*}V_{ud}^L=c^{ZL}_{bd}.
\label{eq:relcoupl1}
\end{equation}
The above equations are satisfied by the left-handed and right-handed components of the couplings, separately. The immediate consequence of these equalities is that the FCNC induced by the $Z$-boson are null if the extended CKM matrix is unitary. We shall analyse the phenomenological implications of this result in more detail in the next section.

\noindent
The second highest order terms increasing with $t$ are more cumbersome as they are proportional to the different masses of the external fermions involved in the scattering process. These terms are generated by all the three amplitudes contributing to the scattering, therefore they include all the fermion couplings to the $W$, $Z$ and Higgs bosons. After applying the relation between the weak couplings given in Eq.~\ref{eq:relcoupl1}, this contribution reads:
\begin{align*} 
I_t &= {g_W^4\over M_W^4}(-t)\Big [(m_b^2+m_d^2){{(|c^{ZL}_{bd}|^2+|c^{ZR}_{bd}|^2)}\over 4}+m_bm_d{{(c^{ZL}_{bd}c^{ZR*}_{bd}+h.c.)}\over 2}+{{(m_L^2|c^{hL}_{bd}|^2+m_R^2|c^{hR}_{bd}|^2)}\over{16}}\\
& +m_b{{(m_Lc^{ZL}_{bd}c^{hL*}_{bd}+m_Rc^{ZR}_{bd}c^{hR*}_{bd}+h.c.)}\over 8}+m_d{{(m_Lc^{ZR}_{bd}c^{hL*}_{bd}+m_Rc^{ZL}_{bd}c^{hR*}_{bd}+h.c.)}\over 8}\Big ]\numberthis \label{eq:It}
\end{align*}  
where couplings and masses have been defined in Sec. \ref{sec:lagrangian}. The first two terms in the above equation clearly show that the Higgs contribution is needed to restore perturbative unitarity. The two terms purely dependent on the $Z$-boson couplings cannot in fact be cancelled out by themselves. Furthermore, from the quadratic dependence of the first term on the $m_b$ and $m_d$ masses it is clear that, in order to suppress the $I_t$ term, the generic masses $m_{L, R}$ entering the fermion-Higgs coupling must be equal to either $m_b$ or $m_d$. This is expected as the Higgs couplings to fermions are proportional to their masses. The novelty in this case, as compared to the Standard Model, is that there are potential FCNC induced by the Higgs connecting two different fermions characterised by two different masses. A priori, one cannot know whether $m_L$ is equal to $m_d$ or $m_b$ (and consequently $m_R$ equal to $m_b$ or $m_d$). The perturbative unitarity requirement allows one to attribute $m_{L, R}$ the correct fermion mass.

\noindent
There are two possible scenarios where the $I_t$ term is nullified:
\begin{equation}
m_L=m_b, ~~~m_R=m_d, ~~~c^{hL}_{bd}=c^h_{bd}=-2c^{ZL}_{bd}, ~~~c^{hR}_{bd}=c^h_{bd}=-2c^{ZL}_{bd}, ~~~c^{zR}_{bd}=0
\label{eq:scenario1}
\end{equation} 
\begin{equation}
m_L=m_d, ~~~m_R=m_b, ~~~c^{hL}_{bd}=c^h_{bd}=-2c^{ZR}_{bd}, ~~~c^{hR}_{bd}=c^h_{bd}=-2c^{ZR}_{bd}, ~~~c^{ZL}_{bd}=0.
\label{eq:scenario2}
\end{equation} 

\noindent
In principle, the above two scenarios could be equally realised. However, if we take as limit case the process $W^+b\rightarrow W^+b$, the latter scenario will be excluded. In fact, for that setup $c^{ZL}_{bb}=-0.5\sum_u|V_{bu}^L|^2\le -0.5$ taking into account the experimental limits on the SM-like positive definite terms and considering the additional contribution of the hypothetical $|V_{bT}^L|^2$ coming from the new heavy $T$-VLQ. This result is clearly not compatible with the zero value of the left-handed $Z$-boson coupling to the bottom-quarks provided within the scenario in Eq.~\ref{eq:scenario2}. One is therefore left with the mass and coupling configuration given in Eq.~\ref{eq:scenario1}.
There, the condition $c^{ZR}_{bd}=0$ applied to the process with one incoming and one outgoing bottom-quark brings a further constraint. From the explicit expression $c^{ZR}_{bb}=-0.5\sum_{u^\prime} |V_{bu^\prime}^R|^2$ with $u^\prime =u, c, t, T$, one can evince that all the entries $V_{bu^\prime}^R$ of the right-handed CKM matrix, extended by the addition of one or more VLQs, are null. 

\noindent
An important advantage of the method we have used in this investigation is that, due to its generality, the amplitudes that have been calculated for the process $W^+d\rightarrow W^+b$ can be interpreted through a number of different frameworks, in each case deriving
unique relations between the couplings. Just for illustrative purposes, so far we have considered interactions involving an initial down quark state and a final bottom quark state. If we so wished, however, the calculated matrix elements could easily describe interactions between any other two types of down-type quark, including the new heavy VLQs,  by simply relabelling the relevant mass terms and the elements of the extended CKM matrix. We could even use it to describe transitions between up-type quarks and/or corresponding VLQs, by replacing any term containing a sum over the up-type quarks, with a sum over down-type quarks instead. Taking into account all possible processes, the net outcome is that all the entries of the right-handed CKM matrix, extended by the addition of one or more VLQs, are null: $V_{qq^\prime}^R=0$ with $q = d, s, b, B$ and $q^\prime = u, c, t, T$.


\noindent
Within this scenario, the last term of the full amplitude square growing with $E_{CM}^2$ via the Mandelstam variables $u$ and $t$ is cancelled out as well. This term depends on the mass of the up-type quarks appearing in the first graph of Fig.~\ref{fig:FD_Sam}; it can be written as:
\begin{align*} 
I_{mu} &= {g_W^4\over M_W^4}\Big\{\Big (-{t\over 4}\Big )\Big [|\sum_um_uV_{bu}^{R*}V_{ud}^L|^2+|\sum_um_uV_{bu}^{L*}V_{ud}^R|^2\Big ]+{u\over 2}m_d\sum_um_u[V_{bu}^{R*}V_{ud}^Lc^{ZR*}+V_{bu}^{L*}V_{ud}^Rc^{ZL*} + H.c.]\\
& +{u\over 2}m_b\sum_um_u[V_{bu}^{L*}V_{ud}^Rc^{ZR*}+V_{bu}^{R*}V_{ud}^Lc^{ZL*} + H.c.].\numberthis \label{eq:It}
\end{align*}  
Since this contribution is proportional to the entries of the right-handed CKM, it automatically vanishes in the above setup. Only the terms constant or decreasing with energy are left, thus recovering the good behaviour of the scattering amplitude at high energies. 

\subsection{Unitary VLQ model}
\label{sec:uvlq-model}

\noindent
In this section, we summarise the VLQ model preserving perturbative unitarity. The masses and couplings must satisfy the following relations:
\begin{equation}
m_L=m_{q}, ~~~m_R=m_{q^\prime}, ~~~c^{hL}_{qq^\prime}=c^{hR}_{qq^\prime}=-2c^{ZL}_{qq^\prime}=\sum_{q^{\prime\prime}}V_{qq^{\prime\prime}}^{L*}V_{q^{\prime\prime}q^\prime}^L, ~~~V_{qq^{\prime\prime}}^R=0, ~~~c^{ZR}_{qq^\prime}=0
\label{eq:scenario}
\end{equation}
where $q$ is the incoming quark, $q^\prime$ is the outgoing quark and $q^{\prime\prime}$ is the quark mediator of the weak interaction. Independently on their representation, vector-like quarks thus behave in the same manner as the SM-like quarks. Only the left-handed VLQs interact via the weak force, the Lagrangian being
\begin{equation}
{\mathcal{L}}_W = -{g_W\over\sqrt{2}}\bar q\gamma^\mu V^L_{qq^\prime}P_L q^\prime W^+_\mu + H.c.
\end{equation}

\noindent
Moreover, if the extended CKM matrix is unitary like in the Standard Model, FCNC are not allowed. Alternatively, the FCNC induced by the $Z$-boson is purely left-handed as the weak charged current is. The weak neutral current Lagrangian is expressed as:
\begin{equation}
{\mathcal{L}}_Z = -{g_W\over{c_W}}\bar q\gamma^\mu \left (\pm{1\over 2}V^{L*}_{qq^{\prime\prime}}V^L_{q^{\prime\prime}q^\prime} P_L - 2Q_qs^2_W\delta_{qq^\prime} \right )q^\prime Z_\mu + H.c.
\end{equation}
where the plus(minus) sign is for down-type(up-type) quarks. 
On the contrary, the FCNC generated by the Higgs has both a left-handed and a right-handed part proportional to the outgoing and incoming quark mass, respectively. The corresponding Lagrangian is given by:
\begin{align*}
{\mathcal{L}}_h &= {g_W\over{2M_W}}\bar q\gamma^\mu V^{L*}_{qq^{\prime\prime}}V^L_{q^{\prime\prime}q^\prime}(m_bP_L + m_d P_R)q^\prime H + H.c. \\
&= {g_W\over{4M_W}}\bar q\gamma^\mu V^{L*}_{qq^{\prime\prime}}V^L_{q^{\prime\prime}q^\prime}((m_b+m_d) + (m_d-m_b)\gamma_5)q^\prime H + H.c.\numberthis \label{eq:lagrangianH}
\end{align*}
Owing to the fermion mass splitting, the Higgs boson acquires a pseudo-scalar component.

\noindent
As noted earlier, the general nature of our approach means that the VLQ model in Eq.~\ref{eq:scenario} can incorporate the new heavy VLQs in singlet or doublet representations, and no assumption has been made about the level of mixing with the SM quarks. This model agrees with the one proposed in Ref.~\cite{Aguilar-Saavedra:2009xmz, Aguilar-Saavedra:2013qpa}, which can be considered as a sub-case of our setup. There, in fact, the $B$-singlet and $T$-singlet representations only are considered under the assumption that the extra VLQs would mix with the third generation SM quarks uniquely. Our result reinforces this model, explicitly showing that the structure of its couplings is dictated by the fundamental requirement of perturbative unitarity. In addition, our findings generalise such a structure to the standard doublet representation, without any conditions on the mixing. 

\begin{figure}[t!]
\begin{center}
\includegraphics[width=0.7\textwidth]{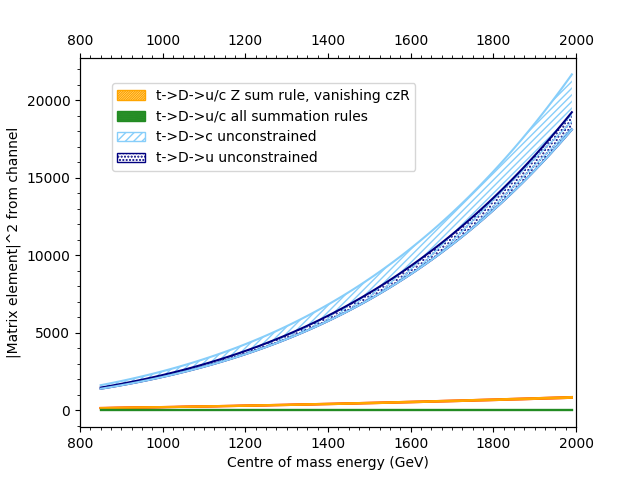}
\caption{Matrix element square for the processes $W^- t\rightarrow W^- u$ and $W^- t\rightarrow W^- c$ versus the CoM energy. We assume $V_{qq^\prime}^R=0$ with $q(q^\prime )$ any up-type (down-type) quarks, including the VLQs, and $c^{ZR}=0$. The blue (cien) contour refers to the scattering amplitude squared of the channel with an outgoing up-quark (charm-quark) when no relations between couplings are imposed. The orange solid line represents the matrix element squared of the two overlapping processes when the relation in Eq.~\ref{eq:relcoupl1} between the $W$- and $Z$-boson couplings to quarks is imposed. The green solid line shows the amplitude square for the two processes when all the relations in Eq.~\ref{eq:scenario} between the $W$-, $Z$- and Higgs-boson couplings to quarks are implemented.}
\label{fig:energiafcnc}
\end{center}
\end{figure}

\noindent
Before discussing the phenomenological implications of our outcomes, we conclude this section with a graphic representation of the behaviour of the amplitude squared corresponding to two illustrative scattering processes, $W^-t\rightarrow W^-u$ and $W^-t\rightarrow W^-c$, as a function of the CoM energy. The choice of these processes, specifically, is motivated by the presence of experimental bounds on the strength of the FCNC inducing both the Higgs-up-top and Higgs-charm-top interactions. In line with our findings and for ease of calculation, we assume $V_{qq^\prime}^R=0$ with $q(q^\prime )$ any up-type (down-type) quarks, including the VLQs, and $c^{ZR}=0$. The blue (cien) contour refers to the scattering amplitude squared of the channel with an outgoing up-quark (charm-quark) when no relations between couplings are imposed. The growth with the CoM energy to the power four, coming from the dominant $us$-term, is clearly displayed. The orange solid line represents the matrix element squared of the two processes, which overlap, when the relation in Eq.~\ref{eq:relcoupl1} between the $W$- and $Z$-boson couplings to quarks is imposed. In this case, the $us$-term is cancelled out and the sub-dominant terms, proportional to the $u$ and $t$ Mandelstam variables, lead the growth of the matrix element squared with the C.o.M. energy to the power two. The green solid line shows the amplitude square for the two processes when all the relations in Eq.~\ref{eq:scenario} between the $W$-, $Z$- and Higgs-boson couplings to quarks are implemented. This is when both the dominant and sub-dominant terms in the CoM energy are washed out so that the amplitude square is given by a constant plus terms decreasing with energy.

\noindent
We will now turn our attention to the phenomenological consequences of imposing the fulfilment of perturbative unitarity within VLQs models.

\section{Phenomenology of the unitary VLQ model}
\label{sec:pheno}

\noindent
Flavour violating couplings between quarks and neutral $Z$ and Higgs bosons can be generated in presence of either extra neutral gauge bosons or new heavy vector-like quarks 
with $+2/3$ and $-1/3$ electric charges. 
Recently, extensive analyses of flavour violation in the presence of new heavy VLQs have been presented \cite{Bobeth:2016llm}
Up to now, the presence of such VLQs has been considered essential for generating flavour violating couplings to SM up- and down-quarks, respectively. Our results show that FCNC are instead correlated to the unitarity of the extended CKM and therefore depend on the type of group representation the VLQs are described by. In the following, we analyse the various representations individually.

\subsection{T-singlet}

\noindent
We consider the SM with the minimal extension of one up-type iso-singlet vector-like quark, $T_{L,R}$, with fractional electric charge $Q=+2/3$. In the flavour basis, the mass terms in the Lagrangian read as:
\begin{equation}
{\mathcal{L}}_m = (\bar u_L ~\bar T_L)M_u \Big (\begin{matrix} u_R\\ T_R\end{matrix}\Big ) + \bar d_LM_dd_R + H.c.
\end{equation}
where $u_L, d_L$ are represented by the SM quark doublets and $u_R, d_R$ by the SM quark singlets. The two mass matrices $M_u$ and $M_d$ can be diagonalised via $4\times 4$ and $3\times 3$ bi-unitary transformations, respectively, as
\begin{equation}
D_u=V_L^{u\dag}M_uV_R^u~~~~~~~{\rm and}~~~~~~~D_d=V_L^{d\dag}M_dV_R^d
\end{equation}
with $D_u=diag(m_u, m_c, m_t, m_T)$, where $m_T$ is the mass of the new heavy up-type VLQ, and $D_d=diag(m_d, m_s, m_b)$. Like in the SM, going from the flavour basis to the mass eigenstates basis, the charged current Lagrangian becomes:
\begin{equation}
{\mathcal{L}}_W = -{g_w\over\sqrt{2}}(\bar u_L ~\bar T_L)V_{CKM}^{ext}\gamma^\mu d_LW^+_\mu + H.c.
\end{equation}
where the mixing matrix, $V_{CKM}^{ext} = V^{u\dag}_LV^d_L$, is now an extended $4\times 3$ matrix 
\begin{equation*}
V_{CKM}^{ext} =
\begin{pmatrix}
V_{ud} & V_{us} & V_{ub} \\
V_{cd} & V_{cs} & V_{cb} \\
V_{td} & V_{ts} & V_{tb} \\
V_{Td} & V_{Ts} & V_{Tb}
\end{pmatrix}
\end{equation*}
whose first three rows play the role of the $3\times 3$ CKM mixing matrix. This latter sub-matrix is not unitary anymore, unless there is no mixing of the SM quarks with the new heavy VLQ. Neither the enlarged $V_{CKM}^{ext}$ is unitary, a priori. Therefore, the appearance of tree-level FCNC mediated by the $Z$ and Higgs bosons is a natural consequence of the $T_{L,R}$ existence. In the physical (or mass) basis, the vectorial neutral current Lagrangian becomes:
\begin{equation}
{\mathcal{L}}_Z = -{g_W\over c_w}\big [{1\over 2}(\bar u_L ~\bar T_L)V_{CKM}^{ext} V_{CKM}^{ext\dag}\gamma^\mu \Big (\begin{matrix} u_L\\ T_L\end{matrix}\Big ) -{1\over 2}\bar d_L V_{CKM}^{ext\dag} V_{CKM}^{ext}\gamma^\mu d_L -{2\over 3}s^2_w(\bar u_i\gamma^\mu u_i + \bar T\gamma^\mu T) + {1\over 3}(\bar d_i\gamma^\mu d_i)\big ]Z_\mu + H.c.
\end{equation}
where the $4\times 4$ $V_{CKM}^{ext} V_{CKM}^{ext\dag}$ and the $3\times 3$ $V_{CKM}^{ext\dag} V_{CKM}^{ext}$ matrices are not diagonal and can induce FCNC in the quark sector. An analogous result arises for the Higgs-mediated FCNC. In this scenario, the FCNC connecting two down-type quarks are suppressed compared to the FCNC giving rise to two up-type quarks. This can be better visualised by choosing a weak basis where the down-type quark mass matrix is diagonal and therefore $V_{L,R}^d=I_3$. The phenomenological consequence is that in the down-type quark sector the couplings inducing the FCNC mediated by the $Z$-boson are null, $c^{ZL}_{bd}=c^{ZL}_{bs}=0$, as well as those generating the FCNC mediated by the Higgs, $c^{hL}_{bd}=c^{hL}_{bs}=0$. Therefore, the $b$-quark physics cannot provide useful observables for this type of VLQ model. On the contrary, the rare top decays into a lighter up-type quark plus the associate production of a neutral boson, $Z$ or Higgs, represent a valuable signature. The recent $95\%$ C.L. exclusion bounds coming from the analyses of  the top decays $t\rightarrow Hu$ and $t\rightarrow Hc$ could be used to constrain the model and extract limits not only on the couplings $c^{hL,R}_{tu}, c^{hL,R}_{tc}$ but also on $c^{ZL}_{tu}, c^{ZL}_{tc}$, using Eq.~\ref{eq:scenario}. In order to estimate the size of such couplings as predicted by the $T$-singlet model, one can use the present measurements of the SM-like entries of the CKM matrix and apply Eq.~\ref{eq:scenario}. One obtains
\begin{equation}
c^{ZL}_{tu}=-{1\over 2}\sum_{D=d,s,b}V^L_{tD}V^{*L}_{uD}=(1.05\pm 0.05)\cdot 10^{-4} + {\rm i} (3.2\pm 0.2)\cdot 10^{-3};~~~~~~~~|c^{ZL}_{tu}|=(3.2\pm 0.2)\cdot 10^{-3} 
\end{equation}
\begin{equation}
c^{ZL}_{tc}=-{1\over 2}\sum_{D=d,s,b}V^L_{tD}V^{*L}_{cD}=(3.90\pm 0.40)\cdot 10^{-4}-{\rm i}(3.7\pm 0.2)\cdot 10^{-4};~~~~~~~~|c^{ZL}_{tc}|=(5.4\pm 0.4)\cdot 10^{-4}
\end{equation}       
\begin{equation}
c^{hL,R}_{tu}=\sum_{D=d,s,b}V^L_{tD}V^{*L}_{uD}=(-2.1\pm 0.1)\cdot 10^{-4} - {\rm i} (6.4\pm 0.3)\cdot 10^{-3};~~~~~~~~|c^{hL,R}_{tu}|=(6.4\pm 0.4)\cdot 10^{-3} 
\end{equation}
\begin{equation}
c^{hL,R}_{tc}=\sum_{D=d,s,b}V^L_{tD}V^{*L}_{cD}=(-7.7\pm 0.7)\cdot 10^{-4} +{\rm i}(7.4\pm 0.4)\cdot 10^{-4};~~~~~~~~|c^{hL,R}_{tc}|=(1.1\pm 0.08)\cdot 10^{-3}
\label{eq:higgs_coupl_estimate}
\end{equation} 

\begin{figure}[t!]
\begin{center}
\vskip -1cm
\includegraphics[width=0.9\textwidth]{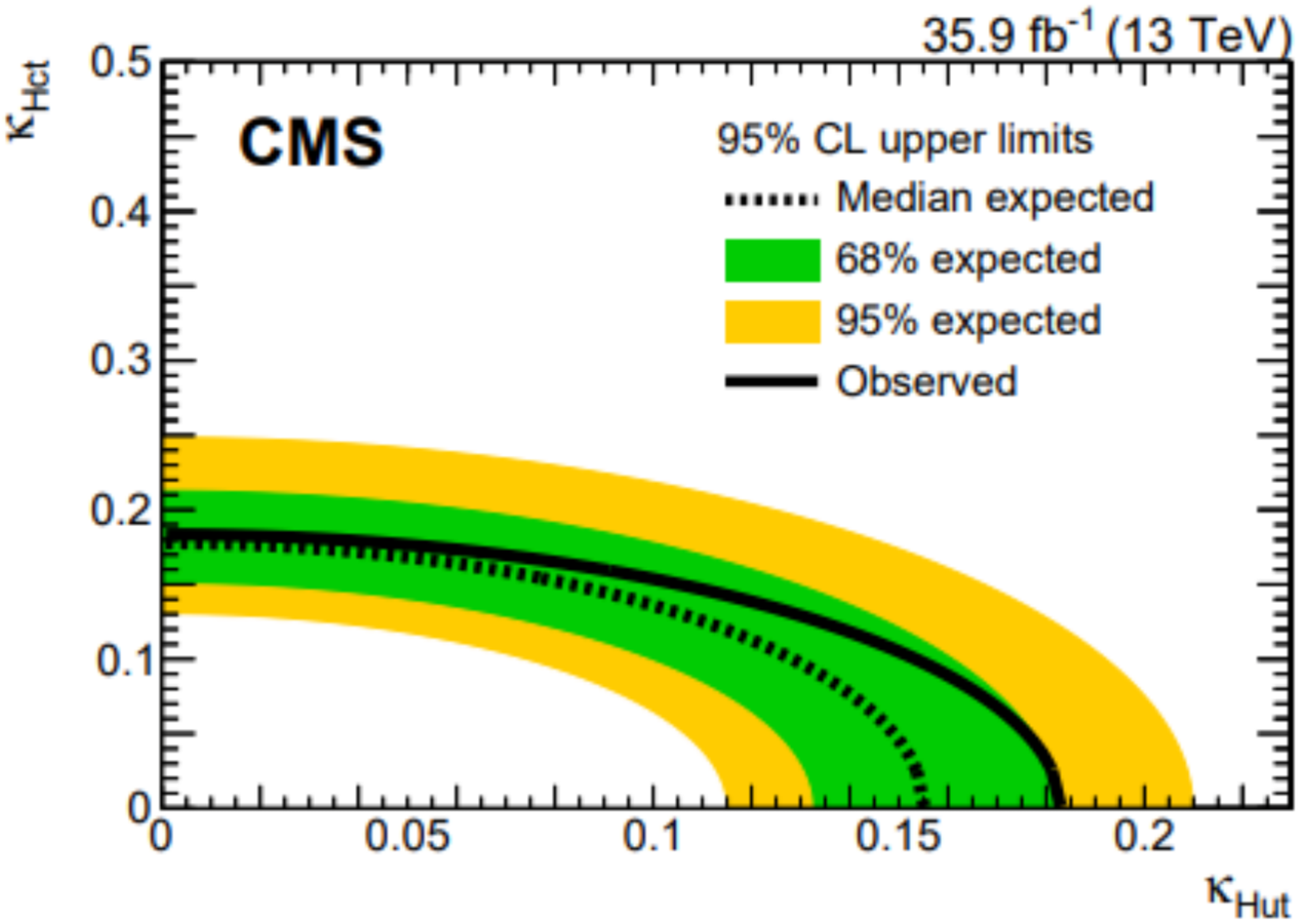}
\vskip -1cm
\caption{Upper limits on Higgs FCNC couplings from CMS experiment (see Ref.~\cite{CMS:2017bhz}).}
\label{fig:CMS-Hct-Hut}
\end{center}
\end{figure}

\begin{figure}[t!]
\begin{center}
\vskip -2cm
\includegraphics[width=0.9\textwidth]{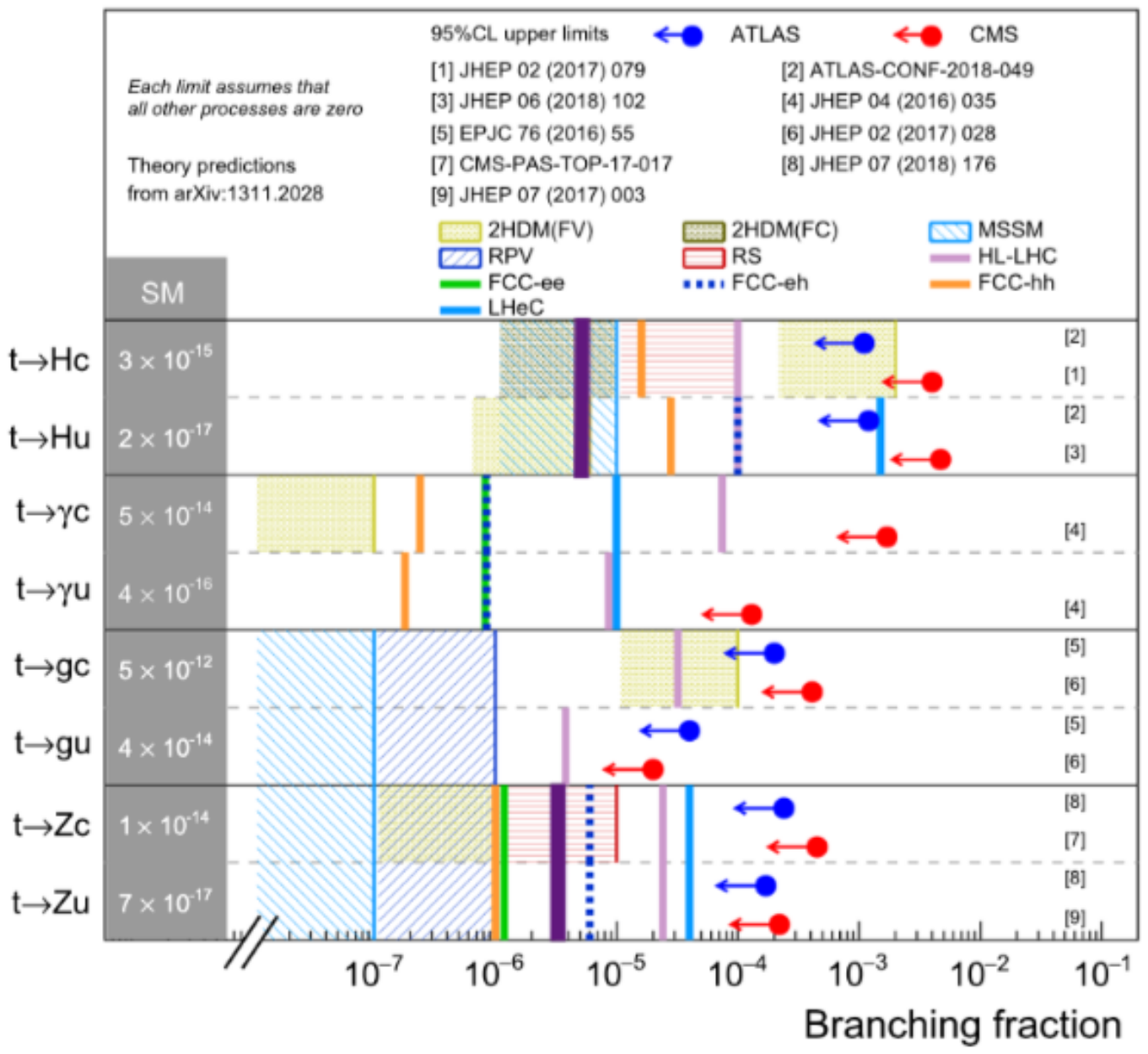}
\vskip -11cm
\caption{Summary of 95$\%$ C.L. limits in the search for FCNC in top production or decays for various future collider options, compared to current LHC limits \cite{FCC:2017aa}. The thick purple lines represent the Branching fractions within the $T$-singlet.}
\label{fig:FigBr}
\end{center}
\end{figure}

\noindent      
These values are more than two orders of magnitude smaller than the couplings probed at the present LHC run. Fig.~\ref{fig:CMS-Hct-Hut} shows the 95$\%$ upper limits obtained by CMS on the neutral couplings inducing the flavour-changing decay of the top-quark into the lighter up-type quarks $u$ and $c$ plus the Higgs. In the CMS analysis of Ref.~\cite{CMS:2017bhz}, the Lagrangian is given by
\begin{equation}
{\mathcal{L}}_h = \sum_{q=u, c}{g_W\over\sqrt{2}}\bar t X_{hqt}(f^L_{hq}P_L + f^R_{hq}P_R)qh + H.c.
\end{equation}
where $X_{hqt}$ is the flavour violating coupling between the top-quark and a lighter up-type quark $q=u, c$, while $f^{L,R}_{hq}$ are complex parameters and $P_{L,R}$ are the projection operators as usual. It is immediate to see that in our notation the FCNC coupling $c_{tq}^{hL}$ can be re-expressed as 
\begin{equation}
c^{hL,R}_{tq}=\sqrt{2}{M_W\over M_t} X_{hqt}. 
\end{equation}
Therefore, considering the scenario given in Eq.~\ref{eq:scenario}, we can recast the parameter that is used to interpret the experimental data and visualise the exclusion bounds as
\begin{equation}
X_{hqt} = {M_t\over{\sqrt{2} M_W}}\sum_{D=d,s,b}V^L_{tD}V^{*L}_{qD} ~~~~~ {\rm with} ~~~~~ q=u, c.
\end{equation}
Inserting the measured values of the CKM matrix entries, as in Eq.~\ref{eq:higgs_coupl_estimate}, we obtain values of the order of permill, which are well below the current sensitivity at the LHC. We can however estimate the branching ratios of the rare decays of the top-quark, via the neutral FCNC mediated by the $Z$-boson and the Higgs, to evaluate the potential of future accelerators in this respect. The partial width of the top decay into a lighter up-type quark and a $Z$-boson is given by
\begin{equation}
\Gamma_Z (T\rightarrow Z q) = {{G_F m_t^3}\over {4\sqrt{2}\pi\cos^2\theta_W}}|c^{ZL}_{tq}|^2\Big (1-{M_Z^2\over m_t^2}\Big )^2\Big (1+2{M_Z^2\over m_t^2}\Big )
\end{equation}
with $q=u, c$ and neglecting their mass. Thus we obtain, $\Gamma_Z (t\rightarrow Z u)=3.6\cdot 10^{-6}$ GeV and $\Gamma_Z (t\rightarrow Z c)=10^{-6}$. Analogously, the partial width of the top decay into a lighter up-type quark and a Higgs boson is given by
\begin{equation}
\Gamma_h (t\rightarrow h q) = {{G_F m_t^3}\over {16\sqrt{2}\pi}}|c^{hL}_{Tt}|^2\Big (1-{M_h^2\over m_t^2}\Big )^2.
\end{equation}
In this case, we get $\Gamma_h (t\rightarrow h u)=7.6\cdot 10^{-6}$ GeV and $\Gamma_h (t\rightarrow h c)=2.2\cdot 10^{-7}$. Taking the top-quark width to be $\Gamma_{top}=1.35$ GeV, the individual branching ratios in the four different channels, assuming that all other processes are zero, are: $Br(t\rightarrow Zu)=2.7\cdot 10^{-6}$, $Br(t\rightarrow Zc)=0.7\cdot 10^{-6}$, $Br(t\rightarrow hu)=5.6\cdot 10^{-6}$ and $Br(t\rightarrow hc)=1.6\cdot 10^{-7}$. From Fig.~\ref{fig:FigBr}, which shows a summary of the 95$\%$ C.L. limits in the search for FCNC in top-quark production and decay for various future colliders, it is clear that FCNC in the top sector generated by the $T$-singlet (see purple lines) could only be observed at the FCC-hh or FCC-ee in the channel $t\rightarrow Zu$.

\noindent 
From the point of view of the direct search for these new heavy particles, the mixing between $T$ and the SM quarks is a priori different from zero. This means that the charged couplings $V_{Td^\prime}^L$ ($d^\prime =d, s, b$) and the neutral couplings $c^{zL}_{Tu^\prime}$ and $c^{h}_{Tu^\prime}$ with ($u^\prime =u, c, t$) could be sizeable. This would allow the new VLQ to be singly produced and afterwards to decay into SM quarks with the associate production of either a vector or a scalar boson: $T\rightarrow W^+d^\prime$, $T\rightarrow Zu^\prime$ and $T\rightarrow Hu^\prime$ with $d^\prime (u^\prime )$ any down-type (up-type) SM quark. 

\noindent
We can estimate the value of such couplings by combining the unitarity condition on the multiplication of any two columns pertaining to the extended CKM matrix with the measurements of the SM-like CKM elements. From the first condition, we get three relations:
\begin{align*}
0 &= \sum_{U=u, c, t}V^L_{dU}V^{L*}_{Ub} + V^L_{dT}V^{L*}_{Tb} = (-3.2\pm 0.75)\cdot 10^{-5}-{\rm i}(6.4\pm 0.2)\cdot 10^{-3} + V^L_{dT}V^{L*}_{Tb} \\
0 &= \sum_{U=u, c, t}V^L_{dU}V^{L*}_{Us} + V^L_{dT}V^{L*}_{Ts} = (-3.4\pm 0.2)\cdot 10^{-4} -{\rm i}(1.4\pm 0.1)\cdot 10^{-4} + V^L_{dT}V^{L*}_{Ts} \\
0 &= \sum_{U=u, c, t}V^L_{sU}V^{L*}_{Ub} + V^L_{sT}V^{L*}_{Tb} = (-7.7\pm 0.7)\cdot 10^{-4}-{\rm i}(7.4\pm 0.4)\cdot 10^{-4} + V^L_{sT}V^{L*}_{Tb} \numberthis \label{eq:VTD-entries}
\end{align*}
where the numerical values are taken from the PDG. Solving the system, the absolute value of the three independent elements of the extended CKM matrix, which represent the mixing between the new $T$ VLQ and the SM down-type quarks, can be estimated to be:
\begin{equation}
V^L_{dT}=0.046; ~~~~~ V^L_{sT}=0.008; ~~~~~ V^L_{bT}=0.14\pm 0.04.
\end{equation}

\noindent
By making use of these estimates, we can now extract values for the neutral FCNC couplings between the $T$ VLQ, an up-type SM quark and either the $Z$ or the Higgs boson. We have in fact
\begin{align*}
c^{hL,R}_{Tu} &= \sum_{D=d,s,b}V^L_{TD}V^{L*}_{Du} \simeq V^L_{Td}V^{L*}_{du} = 0.046\\
c^{hL,R}_{Tc} &= \sum_{D=d,s,b}V^L_{TD}V^{L*}_{Dc} \simeq V^L_{Ts}V^{L*}_{sc} = 0.008\\
c^{hL,R}_{Tt} &= \sum_{D=d,s,b}V^L_{TD}V^{L*}_{Dt} \simeq V^L_{Tb}V^{L*}_{bt} = 0.14
\end{align*}
We are now ready to compute decay widths and branching ratios of the new heavy $T$ VLQ. The decay rate into a $W$-boson and bottom-quark pair increases with the third power of the $T$ VLQ mass 
\begin{align*}
\Gamma_W (T\rightarrow W^+ b) &= {{G_F M_T^3}\over {8\sqrt{2}\pi}}|V_{bT}^L|^2\sqrt{\Big (1-{M_W^2\over M_T^2}-{m_b^2\over M_T^2}\Big )^2-4{{M_W^2m_b^2}\over M_T^4}}\Big [\Big (1-{M_W^2\over M_T^2}+{m_b^2\over M_T^2}\Big )
\Big (1+2{M_W^2\over M_T^2}+{m_b^2\over M_T^2}\Big )-2{m_b^2\over M_T^2}\Big ]\\
&\simeq 175 {\rm MeV} \Big ({M_T\over M_W}\Big )^3|V_{bT}^L|^2.
\end{align*}

\noindent
If we consider as a range of hypothetical masses the interval $1 {\rm TeV}\le M_T\le 5 {\rm TeV}$, the partial width in the charged channel could vary between $7 {\rm GeV}\le \Gamma_W\le 837 {\rm GeV}$. 
As to the neutral channel, the decay rate into a $Z$-boson and top-quark pair is given by
\begin{align*}
\Gamma_Z (T\rightarrow Z t) &= {{G_F M_T^3}\over {4\sqrt{2}\pi\cos^2\theta_W}}|c^{ZL}_{Tt}|^2\sqrt{\Big (1-{M_Z^2\over M_T^2}-{m_t^2\over M_T^2}\Big )^2-4{{M_Z^2m_t^2}\over M_T^4}}\Big [\Big (1-{M_Z^2\over M_T^2}+{m_t^2\over M_T^2}\Big )
\Big (1+2{M_Z^2\over M_T^2}+{m_t^2\over M_T^2}\Big )-2{m_t^2\over M_T^2}\Big ]\\
&\simeq 443 {\rm MeV} \Big ({M_T\over M_W}\Big )^3|c^{ZL}_{Tt}|^2.
\end{align*}
For the same mass range as above, we get a partial width which is three orders of magnitude bigger than the previous one: $4.2 {\rm GeV}\le \Gamma_Z\le 530 {\rm GeV}$. Analogously, for the decay rate into a Higgs boson and top-quark pair we have
\begin{align*}
\Gamma_h (T\rightarrow h t) &= {{G_F M_T^3}\over {16\sqrt{2}\pi}}|c^{hL}_{Tt}|^2\sqrt{\Big (1-{M_h^2\over M_T^2}-{m_t^2\over M_T^2}\Big )^2-4{{M_h^2m_t^2}\over M_T^4}}\Big [\Big (1-{M_h^2\over M_T^2}+{m_t^2\over M_T^2}\Big )
\Big (1+{m_t^2\over M_T^2}\Big )+{m_t^2\over M_T^2}\Big ]\\
&\simeq 85 {\rm MeV} \Big ({M_T\over M_W}\Big )^3|c^{hL}_{Tt}|^2.
\end{align*}
For the same mass range as above, we get a partial width of the same order as the previous one: $3.3 {\rm GeV}\le \Gamma_h\le 407 {\rm GeV}$. The total width can be easily computed for the same range of masses: $15 {\rm GeV}\le \Gamma_{TOT}\le 1774 {\rm GeV}$. This means that the $T$-singlet VLQ would appear as a narrow-mid resonance. For masses $M_T\le 2650$ GeV, the $T$-singlet would have ${\Gamma_{TOT}\over M_T}\le 10\%$ and could be analysed in narrow width approximation. For higher masses, $2.7\le M_T\le 5$ TeV, the new $T$ VLQ would be broader, having a width over mass ratio $10\%\le {\Gamma_{TOT}\over M_T}\le 35\%$. The signal could manifest itself as a large shoulder spread over the SM background.  Novel strategies, specifically tailored for wide resonances, should be applied. Moreover, given the very large width of the predicted $T$-singlet, heavy interferences with the SM contribution to the same final state could take place, a priori. Therefore, finite width and interference effects should be both taken into consideration when searching for these signals in the mid-large mass regime.

\noindent
For sake of completeness, at fixed mass, we can also calculate the expected branching ratio of the three individual channels: 
\begin{equation}
Br(T\rightarrow Wb)= 47\%; ~~~ Br(T\rightarrow Zt)= 30\%; ~~~ Br(T\rightarrow ht)= 23\%.
\end{equation}

\subsection{B-singlet}

\noindent
Analogously to what done in the previous section, we now consider the SM with the minimal extension of one down-type iso-singlet vector-like quark, $B_{L,R}$, with fractional electric charge $Q=-1/3$. In the flavour basis, the mass terms in the Lagrangian read as:
\begin{equation}
{\mathcal{L}}_m = (\bar d_L ~\bar B_L)M_d \Big (\begin{matrix} d_R\\ B_R\end{matrix}\Big ) + \bar u_LM_uu_R + H.c.
\end{equation}
where $u_L, d_L$ are represented by the SM quark doublets and $u_R, d_R$ by the SM quark singlets. The two mass matrices $M_u$ and $M_d$ can be diagonalised via $3\times 3$ and $4\times 4$ bi-unitary transformations, respectively, as
\begin{equation}
D_u=V_L^{u\dag}M_uV_R^u~~~~~~~{\rm and}~~~~~~~D_d=V_L^{d\dag}M_dV_R^d
\end{equation}
with $D_d=diag(m_d, m_s, m_b, m_B)$, where $m_B$ is the mass of the new heavy down-type VLQ, and $D_u=diag(m_u, m_c, m_t)$. Like in the SM, going from the flavour basis to the mass eigenstates basis, the charged current Lagrangian becomes:
\begin{equation}
{\mathcal{L}}_W = -{g_w\over\sqrt{2}}(\bar d_L ~\bar B_L)V_{CKM}^{ext}\gamma^\mu u_LW^-_\mu + H.c.
\end{equation}
where the mixing matrix, $V_{CKM}^{ext} = V^{d\dag}_LV^u_L$, is now an extended $3\times 4$ matrix
\begin{equation*}
V_{CKM}^{ext} =
\begin{pmatrix}
V_{ud} & V_{us} & V_{ub} & V_{uB}\\
V_{cd} & V_{cs} & V_{cb} & V_{cB}\\
V_{td} & V_{ts} & V_{tb} & V_{tB}
\end{pmatrix}
\end{equation*}
whose first three columns play the role of the $3\times 3$ CKM mixing matrix.
This latter sub-matrix is not unitary anymore, unless there is no mixing of the SM quarks with the new heavy VLQ. Neither the enlarged $V_{CKM}^{ext}$ is unitary, a priori. Therefore, as for the T-singlet above, the appearance of tree-level FCNC mediated by the $Z$ and Higgs bosons is a natural consequence of the $B_{L,R}$ existence as well. In the physical (or mass) basis, the vectorial neutral current Lagrangian becomes:
\begin{equation}
{\mathcal{L}}_Z = -{g_w\over c_w}\big [{1\over 2}(\bar d_L ~\bar B_L)V_{CKM}^{ext} V_{CKM}^{ext\dag}\gamma^\mu \Big (\begin{matrix} d_L\\ B_L\end{matrix}\Big ) -{1\over 2}\bar u_L V_{CKM}^{ext\dag} V_{CKM}^{ext}\gamma^\mu u_L -{2\over 3}s^2_w(\bar u_i\gamma^\mu u_i + \bar T\gamma^\mu T) + {1\over 3}(\bar d_i\gamma^\mu d_i)\big ]Z_\mu + H.c.
\end{equation}
where the $4\times 4$ $V_{CKM}^{ext} V_{CKM}^{ext\dag}$ and the $3\times 3$ $V_{CKM}^{ext\dag} V_{CKM}^{ext}$ matrices are not diagonal and can induce FCNC in the quark sector. An analogous result arises for the Higgs-mediated FCNC. In this scenario, the FCNC connecting two up-type quarks are suppressed compared to the FCNC giving rise to two down-type quarks. This can be better visualised by choosing a weak basis where the up-type quark mass matrix is diagonal and therefore $V_{L,R}^u=I_3$. The phenomenological consequence is that the couplings inducing the FCNC mediated by the $Z$-boson are null, $c^{ZL}_{tu}=c^{ZL}_{tc}=0$, as well as those generating the FCNC mediated by the Higgs, $c^{hL}_{tu}=c^{hL}_{tc}=0$. The rare top decays into a lighter up-type quark plus the associate production of a neutral boson, $Z$ or Higgs, are not a signature for this type of VLQ model. Therefore, the recent exclusion bounds coming from these neutral FCNC couplings cannot be used to constrain the model. One would need experimental analyses on the down-type quark sector to claim evidence or extract bounds on the couplings $c^{ZL}_{bd}, c^{ZL}_{bs}, c^{hL,R}_{bd}, c^{hL,R}_{bs}$.       

\noindent
From the point of view of the direct search of these new heavy particles, the mixing between $B$ and the SM quarks is a priori different from zero. This means that the charged couplings $V_{Bu^\prime}^L$ ($u^\prime =u, c, t$) and the neutral couplings $c^{ZL}_{Bd^\prime}$ and $c^{hL,R}_{Bd^\prime}$ with ($d^\prime =d, s, b$) could be sizeable. This would allow the new VLQ to be singly produced and afterwards to decay into SM quarks with the associate production of either a vector or a scalar boson: $B\rightarrow W^-u^\prime$, $B\rightarrow Zd^\prime$ and $B\rightarrow Hd^\prime$ with $d^\prime (u^\prime )$ any down-type (up-type) SM quark. 

\noindent
We can estimate the value of such couplings by combining the unitarity condition on the multiplication of any two rows pertaining to the extended CKM matrix with the measurements of the SM-like CKM elements. From the first condition, we get three relations:
\begin{align*}
0 &= \sum_{D=d, s, b}V^L_{uD}V^{L*}_{Dc} + V^L_{uB}V^{L*}_{Bc} = 6.4\cdot 10^{-5} -{\rm i}1.7\cdot 10^{-4} + V^L_{uB}V^{L*}_{Bc} \\
0 &= \sum_{D=d, s, b}V^L_{uD}V^{L*}_{Dt} + V^L_{uB}V^{L*}_{Bt} = -2.1\cdot 10^{-4}-{\rm i}6.4\cdot 10^{-3}  + V^L_{uB}V^{L*}_{Bt} \\
0 &= \sum_{D=d, s, b}V^L_{cD}V^{L*}_{Dt} + V^L_{cB}V^{L*}_{Bt} = -7.7\cdot 10^{-4}+{\rm i} 7.4\cdot 10^{-4} + V^L_{cB}V^{L*}_{Bt} \numberthis \label{eq:VTD-entries}
\end{align*}
where the numerical values are taken from the PDG. Solving the system, the value of the three independent elements of the extended CKM matrix, which represent the mixing between the new $B$ VLQ and the SM up-type quarks, can be estimated to be:
\begin{equation}
V^L_{uB}=0.03; ~~~~~ V^L_{cB}=0.06; ~~~~~ V^L_{tB}=0.2.
\end{equation}

\noindent
By making use of these estimates, we can now extract values for the neutral FCNC couplings between the $B$ VLQ, a down-type SM quark and either the $Z$ or the Higgs boson. We have in fact
\begin{align*}
c^{hL,R}_{Bd} &= \sum_{U=u,c,t}V^L_{BU}V^{L*}_{Ud} \simeq V^L_{Bu}V^{L*}_{ud} = 0.03\\
c^{hL,R}_{Bs} &= \sum_{U=u,c,t}V^L_{BU}V^{L*}_{Us} \simeq V^L_{Bc}V^{L*}_{cs} = 0.06\\
c^{hL,R}_{Bb} &= \sum_{U=u,c,t}V^L_{BU}V^{L*}_{Ub} \simeq V^L_{Bt}V^{L*}_{tb} = 0.2.
\end{align*}
We are now ready to compute decay widths and branching ratios of the new heavy $T$ VLQ. The decay rate into a $W$-boson and bottom-quark pair increases with the third power of the $T$ VLQ mass 
\begin{align*}
\Gamma_W (B\rightarrow W^+ t) &= {{G_F M_B^3}\over {8\sqrt{2}\pi}}|V_{tB}^L|^2\sqrt{\Big (1-{M_W^2\over M_B^2}-{m_t^2\over M_B^2}\Big )^2-4{{M_W^2m_t^2}\over M_B^4}}\Big [\Big (1-{M_W^2\over M_B^2}+{m_t^2\over M_B^2}\Big )
\Big (1+2{M_W^2\over M_B^2}+{m_t^2\over M_B^2}\Big )-2{m_t^2\over M_B^2}\Big ]\\
&\simeq 175 {\rm MeV} \Big ({M_B\over M_W}\Big )^3|V_{tB}^L|^2.
\end{align*}
If we consider as a range of hypothetical masses the interval $1 {\rm TeV}\le M_B\le 5 {\rm TeV}$, the partial width in the charged channel could vary between $14 {\rm GeV}\le \Gamma_W\le 1.71 {\rm TeV}$. 
As to the neutral channel, the decay rate into a $Z$-boson and top-quark pair is given by
\begin{align*}
\Gamma_Z (B\rightarrow Z b) &= {{G_F M_B^3}\over {4\sqrt{2}\pi\cos^2\theta_W}}|c^{ZL}_{Bb}|^2\sqrt{\Big (1-{M_Z^2\over M_B^2}-{m_b^2\over M_B^2}\Big )^2-4{{M_Z^2m_b^2}\over M_B^4}}\Big [\Big (1-{M_Z^2\over M_B^2}+{m_b^2\over M_B^2}\Big )
\Big (1+2{M_Z^2\over M_B^2}+{m_b^2\over M_B^2}\Big )-2{m_b^2\over M_B^2}\Big ]\\
&\simeq 443 {\rm MeV} \Big ({M_B\over M_W}\Big )^3|c^{ZL}_{Bb}|^2.
\end{align*}
For the same mass range as above, we get a partial width which is three orders of magnitude bigger than the previous one: $9 {\rm GeV}\le \Gamma_Z\le 1.08 {\rm TeV}$. Analogously, for the decay rate into a Higgs boson and top-quark pair we have
\begin{align*}
\Gamma_h (B\rightarrow h b) &= {{G_F M_B^3}\over {16\sqrt{2}\pi}}|c^{hL}_{Bb}|^2\sqrt{\Big (1-{M_h^2\over M_B^2}-{m_b^2\over M_B^2}\Big )^2-4{{M_h^2m_b^2}\over M_B^4}}\Big [\Big (1-{M_h^2\over M_B^2}+{m_b^2\over M_B^2}\Big )
\Big (1+{m_b^2\over M_B^2}\Big )+{m_b^2\over M_B^2}\Big ]\\
&\simeq 85 {\rm MeV} \Big ({M_B\over M_W}\Big )^3|c^{hL}_{Bb}|^2.
\end{align*}
For the same mass range as above, we get a partial width of the same order as the previous one: $7 {\rm GeV}\le \Gamma_h\le 0.8 {\rm TeV}$. The total width can be easily computed for the same range of masses: $30 {\rm GeV}\le \Gamma_{TOT}\le 3.6 {\rm TeV}$. This means that the $B$-singlet VLQ would appear either as a narrow or a wide resonance, depending on its mass. For masses $M_B\le 1900$ GeV, the $B$-singlet would have ${\Gamma_{TOT}\over M_B}\le 10\%$ and could be analysed in narrow width approximation. For higher masses, $1.9\le M_T\le 5$ TeV, the new $B$ VLQ would be much broader, having a width over mass ratio $10\%\le {\Gamma_{TOT}\over M_B}\le 70\%$. The signal could manifest itself as a large shoulder or even worse as an excess of events spread over the SM background. Present searches, performed under the narrow width approximation, might have missed these new particles. Novel strategies, specifically tailored for wide resonances, should be applied. Moreover, given the very large width of the predicted $B$-singlet, heavy interferences with the SM contribution to the same final state could take place, a priori. Therefore, finite width and interference effects should be both taken into consideration when searching for these signals in the mid-large mass regime.

\noindent
For sake of completeness, at fixed mass, we can also calculate the expected branching ratio of the three individual channels: 
\begin{equation}
Br(B\rightarrow Wt)= 48\%; ~~~ Br(B\rightarrow Zb)= 30\%; ~~~ Br(B\rightarrow hb)= 22\%.
\end{equation}

\subsection{(T, B)-doublet}

\noindent
We consider the SM with the minimal extension of one doublet containing one up-type and one down-type vector-like quark, $T_{L,R}$ and $B_{L,R}$, with fractional electric charge $Q=+2/3$ and $Q=-1/3$, respectively. This doublet can be seen as an heavier copy of the SM one. In the flavour basis, the mass terms in the Lagrangian read as:
\begin{equation}
{\mathcal{L}}_m = (\bar u_L ~\bar T_L)M_u \Big (\begin{matrix} u_R\\ T_R\end{matrix}\Big ) + (\bar d_L ~\bar B_L)M_d \Big (\begin{matrix} d_R\\ D_R\end{matrix}\Big ) + H.c.
\end{equation}
where $u_L, d_L$ are represented by the SM quark doublets and $u_R, d_R$ by the SM quark singlets. The two mass matrices $M_u$ and $M_d$ can be diagonalised via $4\times 4$ bi-unitary transformations as
\begin{equation}
D_u=V_L^{u\dag}M_uV_R^u~~~~~~~{\rm and}~~~~~~~D_d=V_L^{d\dag}M_dV_R^d
\end{equation}
with $D_u=diag(m_u, m_c, m_t, m_T)$ and $D_d=diag(m_d, m_s, m_b, m_B)$, where $m_T$ and $m_B$ are the masses of the two new heavy up-type and down-type VLQs. Like in the SM, going from the flavour basis to the mass eigenstates basis, the charged current Lagrangian becomes:
\begin{equation}
{\mathcal{L}}_W = -{g_w\over\sqrt{2}}(\bar u_L ~\bar T_L)V_{CKM}^{ext}\gamma^\mu \Big (\begin{matrix} d_L\\ B_L\end{matrix}\Big )W^+_\mu + H.c.
\end{equation}
where the mixing matrix, $V_{CKM}^{ext} = V^{u\dag}_LV^d_L$, is now an extended $4\times 4$ matrix 
\begin{equation*}
V_{CKM}^{ext} =
\begin{pmatrix}
V_{ud} & V_{us} & V_{ub} & V_{uB}\\
V_{cd} & V_{cs} & V_{cb} & V_{cB}\\
V_{td} & V_{ts} & V_{tb} & V_{tB}\\
V_{Td} & V_{Ts} & V_{Tb} & V_{TB}
\end{pmatrix}
\end{equation*}
including the $3\times 3$ CKM mixing matrix in the upper left corner. This latter sub-matrix is not unitary anymore, unless there is no mixing of the SM quarks with the new heavy VLQ. Oppositely to the T-singlet and B-singlet cases, now the enlarged $V_{CKM}^{ext}$ is unitary by construction (assuming there is no new physics beyond the VLQs). 

\noindent
We now proceed by examining a few specific assumptions that may affect the predictions which are made by our unitary VLQ model. As noted earlier, the general nature of our approach means that this model can incorporate vector-like quarks in singlet or doublet representations, and no assumption has been made about the level of mixing with the standard model quarks. Therefore, we are free to analyse each case individually.

\noindent
If we decide to permit mixing with all generations of quarks, all the entries of the extended CKM matrix are non zero, a priori. From the point of view of the new heavy particles search, the mixing between the new heavy $T$ and $B$ VLQs with the SM quarks may in principle be different from zero as well. This means that the charged couplings $V_{Td^\prime}^L$ ($d^\prime =d, s, b$) and $V_{Bu^\prime}^L$ ($u^\prime =u, c, t$) could be sizeable. On the contrary, as the extended $4\times 4$ CKM matrix is unitary, the tree-level FCNC mediated by the $Z$ and Higgs bosons are absent, owing to the relation between the charged and neutral couplings shown in Eq.~\ref{eq:scenario}. 

\noindent
In the physical (or mass) basis, the vectorial neutral current Lagrangian becomes:
\begin{equation}
{\mathcal{L}}_Z = -{g_w\over c_w}\big [{1\over 2}(\bar u_L ~\bar T_L)\gamma^\mu \Big (\begin{matrix} u_L\\ T_L\end{matrix}\Big ) -{1\over 2}\bar d_L \gamma^\mu d_L -{2\over 3}s^2_w(\bar u_i\gamma^\mu u_i + \bar T\gamma^\mu T) + {1\over 3}(\bar d_i\gamma^\mu d_i)\big ]Z_\mu + H.c.
\end{equation}
where only same flavour quarks can couple. An analogous result arises for the Higgs-mediated neutral current. This implies that the neutral couplings $c^{zL}_{Tu^\prime}$,  $c^{h}_{Tu^\prime}$ with ($u^\prime =u, c, t$) and $c^{zL}_{Bd^\prime}$,  $c^{h}_{Bd^\prime}$ with ($d^\prime =d, s, b$) are zero. This would allow the new VLQ to be singly produced and afterwards to decay into SM quarks with the associate production of a $W$-boson only: $T\rightarrow W^+d^\prime$ with $d^\prime$ any down-type SM quark and $B\rightarrow W^-u^\prime$ with $u^\prime$ any up-type SM quark. The neutral current single production and decay would not be allowed instead, for the above-mentioned reasons. Therefore, the search for these new particles should be different from what performed at the moment. Only the charged channel is indeed open for the single production and the sub-sequent decay of these VLQs.

\noindent
By exploiting the six unitarity conditions on the product of any two of the first three columns and any two of the first three rows, one can estimate the magnitude of the following matrix elements:
\begin{equation}
V^L_{dT}=0.046; ~~~~~ V^L_{sT}=0.008; ~~~~~ V^L_{bT}=0.14\pm 0.04
\end{equation}
\begin{equation}
V^L_{uB}=0.03; ~~~~~ V^L_{cB}=0.06; ~~~~~ V^L_{tB}=0.2\pm 0.04.
\end{equation}
With these values, the width of the $T$ VLQ
\begin{align*}
\Gamma (T\rightarrow W^+ b) &= {{G_F M_T^3}\over {8\sqrt{2}\pi}}|V_{bT}^L|^2\sqrt{\Big (1-{M_W^2\over M_T^2}-{m_b^2\over M_T^2}\Big )^2-4{{M_W^2m_b^2}\over M_T^4}}\Big [\Big (1-{M_W^2\over M_T^2}+{m_b^2\over M_T^2}\Big )
\Big (1+2{M_W^2\over M_T^2}+{m_b^2\over M_T^2}\Big )-2{m_b^2\over M_T^2}\Big ]\\
&\simeq 175 {\rm MeV} \Big ({M_T\over M_W}\Big )^3|V_{bT}^L|^2
\end{align*}
can vary between $7 {\rm GeV}\le \Gamma_W\le 837 {\rm GeV}$, if we consider as a range of hypothetical masses the interval $1 {\rm TeV}\le M_T\le 5 {\rm TeV}$. The width of the $B$ VLQ would be instead
\begin{align*}
\Gamma (B\rightarrow W^+ t) &= {{G_F M_B^3}\over {8\sqrt{2}\pi}}|V_{tB}^L|^2\sqrt{\Big (1-{M_W^2\over M_B^2}-{m_t^2\over M_B^2}\Big )^2-4{{M_W^2m_t^2}\over M_B^4}}\Big [\Big (1-{M_W^2\over M_B^2}+{m_t^2\over M_B^2}\Big )
\Big (1+2{M_W^2\over M_B^2}+{m_t^2\over M_B^2}\Big )-2{m_t^2\over M_B^2}\Big ]\\
&\simeq 175 {\rm MeV} \Big ({M_B\over M_W}\Big )^3|V_{tB}^L|^2
\end{align*}
and it could vary between $14 {\rm GeV}\le \Gamma_W\le 1.71 {\rm TeV}$, if we consider as a range of hypothetical masses the interval $1 {\rm TeV}\le M_B\le 5 {\rm TeV}$. These values show that the two new VLQs in $(T, B)$-doublet would typically appear as narrow-mid  resonances. The $T$ VLQ would have ${\Gamma_{TOT}\over M_T}\le 10\%$ for masses $M_T\le 3850$ GeV and could be analysed in narrow width approximation. Only in the large mass domain, $3.9\le M_T\le 5$ TeV, the new $T$ VLQ would be slightly broader, having a width over mass ratio $10\%\le {\Gamma_{TOT}\over M_B}\le 17\%$. The $B$ VLQ would have ${\Gamma_{TOT}\over M_T}\le 10\%$ for masses $M_B\le 2700$ GeV and there it could be analysed in narrow width approximation. In the mid-large mass domain, $2.7\le M_T\le 5$ TeV, the new $B$ VLQ would be broader, having a width over mass ratio $10\%\le {\Gamma_{TOT}\over M_B}\le 34\%$.   

\noindent
If we assume the existence of new physics beyond the VLQs, the unitarity condition of the $4\times 4$ CKM matrix could be released. This matrix could in fact be considered as a sub-matrix belonging to a larger unitary matrix. In this event, both charged currents and  FCNCs could be present across all generations of up-type and down-type quarks. The new heavy VLQs could then be singly-produced and subsequently decay into SM quarks plus the associate production of a $W$, $Z$ or Higgs boson.
  
\noindent
A different assumption, largely present in the literature, is that VLQs would couple only to the third generation of SM quarks. 
If we stipulate that the $4\times 4$ extended matrix $V=V_{CKM}^{ext}$ is unitary, then it must satisfy 
\begin{equation}
\sum_k V_{ik}V^*_{kj}=0
\end{equation}
for any different $i$ and $j$. If we set $i=T$ (or $i=B$ indifferently), the above-mentioned unitarity condition becomes just 
\begin{equation}
\sum_k V_{Tk}V^*_{kj} = V_{Tb}V^*_{bj} = 0; ~~~~~~~~~~~ \sum_k V_{Bk}V^*_{kj} = V_{Bt}V^*_{tj} = 0
\end{equation}
as the $T(B)$-quark couples only to the $b(t)$-quark. However, In general, the $V^*_{bj}$ and $V^*_{tj}$ entries take demonstrably non-zero values for all other up(down)-type quarks $j$. Thus, for the above condition to hold, it must be the case that 
\begin{equation}
V_{Tb}=0 ~~~~~~~~ {\rm and} ~~~~~~~~ V_{Bt}=0. 
\end{equation}
In this case, no mixing is present neither between the $T$-VLQ and the SM $b$-quark nor between the $B$-VLQ and the SM $t$-quark. This realisation of the VLQ model has troubling phenomenological implications to say the least, as it means that VLQs cannot decay into the lighter SM quarks plus the associated emission of a $W$-boson. Neither they can decay into SM quarks via FCNCs mediated by the $Z$ or the Higgs boson, in virtue of the relation in Eq.~\ref{eq:scenario} and the unitarity of the extended CKM matrix. Along the same reasoning, they cannot be singly produced. Barring some other form of radiative decay, the $T$ and $B$ quarks must therefore be either long-lived or stable particles, calling in to question the predictive capabilities of this particular implementation. Cosmological constraints would indeed forbid the existence of electrically charged stable particles, the VLQs in this case, as source of Dark Matter.

\noindent
If we do not require that the extended CKM matrix is unitary, then this problem does not arise, and we should observe new signals as the $T(B)$-quark couplings with SM ordinary matter are no longer necessarily zero and mixing effects may be observed in interactions involving the SM third-generation quarks. Not only the new heavy VLQs could interact with the third generation SM quarks via the charged weak current, but also the possibility of FCNCs mediated by the $Z$ and the Higgs boson could open up, in principle. The $(T, B)$-doublet scenario, with mixing in the third-generation only, requires therefore to assume 
the existence of new physics beyond the VLQs to be realised. This is consistent with findings by other groups in more model specific approaches.

\section{Conclusions}
\label{sec:conc}
\noindent
In this paper, we have presented a new Vector-like quark model constrained by the requirement of preserving the perturbative unitarity of the theory. We have considered a  general framework, where the SM is extended by the sole presence of VLQs. We have chosen to address, specifically, three group representations: $T$-singlet, $B$-singlet and $(T, B)$-doublet. The first outcome is that perturbative unitarity imposes three main relations between charged and neutral current couplings between quarks (including VLQs) and bosons. These sum rules highly constrain the structure of such couplings and their magnitude. They moreover relate such couplings to the properties of the extended CKM matrix, whether being unitary or not. From this point of view, two are the main consequences. First, the extended CKM matrix is purely left-handed. Second, the SM-like Higgs displays a pseudo- scalar component in its FCNC couplings, which is proportional to the difference in mass of the involved quarks.

\noindent
This result gives rise to novel phenomenological aspects relevant to VLQs, which might have a significant impact on the search strategies at the LHC. For $T$-singlet VLQs, only the up-type quark sector can have FCNCs mediated by the $Z$ and Higgs bosons, therefore $b$-quark physics cannot provide useful observables. On the contrary, the rare top decays into a lighter up-type quark and a neutral boson, $Z$ or Higgs, represent a valuable signature a priori. The branching ratios of the rare top decays within this model could be within the reach of the FCC-ee and FCC-hh. As to the direct search, the $T$-quarks could be either doubly- or singly-produced. They would then consequently decay in all three possible channels, charged and neutral, with branching ratios $Br(T\rightarrow Wb)=47\%, Br(T\rightarrow Zt)=30\%$ and $Br(T\rightarrow Ht)=23\%$. The results of the data analyses obtained at the LHC, assuming 100\% decay into a single channel, should be then appropriately rescaled to extract the exclusion bounds on the mass of the VLQs. Moreover, the search strategy should be tailored according to the $T$-quark mass and width. In the mid-mass range, $M_T\le 2650 {\rm GeV}$, the VLQs would in fact appear as narrow resonances, while for higher masses their width-over-mass ratio could be $35\%$ or more.

\noindent
An analog result holds for the $B$-singlet VLQ. In this case, only the down-type quark sector can have FCNCs mediated by the $Z$ and Higgs bosons, therefore the rare top decays into a lighter up-type quark and a neutral boson, $Z$ or Higgs, cannot provide any useful measurement. The $B$-quarks could instead contribute to $b$-quark physics. As the $T$-quark, also the $B$-quark could be doubly or singly produced and then decay into SM quarks and bosons with similar branching ratios as its $T$ counterpart. A noticeable difference compared to the $T$-quark consists in the width of the $B$-quark. This time, while for $M_T\le 1900 {\rm GeV}$ they would appear as narrow resonances, with increasing the mass their width-over-mass value could reach $70\%$ and more. The signal would manifest itself as a large shoulder or even worse as an excess of events spread over the SM background. 
Present searches, performed under the narrow width approximation, might have missed these new particles.

\noindent
The $(T, B)$-doublet case is instead different in more than one aspect. If the extended CKM is assumed to be unitary, all FCNCs are absent independently on the type of quarks. This would allow the new VLQs to be doubly- or singly-produced and afterwards to decay into SM quarks with the associate production of a $W$-boson, only. Therefore, the search for these new particles should be different from what performed at the moment; only the charged channel should be considered. If in the doublet representation, the $T$ and $B$ quarks are expected to appear as rather narrow resonances, with a width-over-mass ratio not exceeding $35\%$. In a scenario where the extended CKM were not unitary, all decay channels would open up again. Finally, a comment on an assumption largely present in the literature, that is VLQs only couple to the third generation of SM quarks. In this scenario, if we stipulate that the extended CKM is unitary, no mixing would be present between VLQs and SM quarks rendering the VLQs (nearly) stable particles. Cosmological constraints would indeed forbid the existence of electrically charged stable particles as source of Dark Matter, so the model would be ruled out. Relaxing the unitarity condition of the CKM would of course make the scenario viable again. The $(T,B)$-doublet scenario, with mixing in the third-generation only, requires therefore to assume the existence of new physics beyond the VLQs to be realised. 

\noindent
To summarise, the unitary VLQ model constrained by the requirement of perturbative unitarity is more predictable than the more general VLQ models and could be tested at the upcoming LHC run 3.

\section*{Acknowledgements}
\noindent
This work is supported by the Science and Technology Facilities Council, grant number  
ST/L000296/1. EA acknowledges partial financial support through the NExT Institute.

\bibliographystyle{apsrev4-1}
\bibliography{VLQ-1002}

\end{document}